\documentclass[11pt, oneside]{amsart}
\usepackage{geometry, appendix}                
\usepackage{graphicx}
\usepackage{amssymb, amsthm, mathrsfs, bm}
\usepackage{epstopdf}
\usepackage{enumerate}
\usepackage{caption}
\usepackage{subcaption}
\usepackage{color, nicematrix}
\usepackage[colorlinks, allcolors=black, linkcolor=black]{hyperref}
\usepackage{upgreek}
\usepackage{graphicx}
\usepackage{graphics, cancel} 
\usepackage{color}
\usepackage{amsmath, amsthm, slashed}
\usepackage{amssymb}
\usepackage{rotating}
\usepackage{epsfig}
\usepackage{verbatim, longtable}
\usepackage{rotating}
\usepackage{graphicx}
\newcommand{\be}{\begin{equation*}}
\newcommand{\ee}{\end{equation*}}
\newcommand{\ben}[1]{\begin{equation}\label{#1}}
\newcommand{\een}{\end{equation}}
\newcommand{\bea}{\begin{eqnarray}}
\newcommand{\eea}{\end{eqnarray}}
\newcommand{\bean}{\begin{eqnarray*}}
\newcommand{\eean}{\end{eqnarray*}}

\newcommand{\R}{\mathbb{R}}
\newcommand{\A}{\A}

\newcommand{\C}{\mathbb{C}}

\renewcommand{\A}{\mathcal{A}}

\newcommand{\abs}[1]{\left|#1 \right|} 
\newcommand{\norm}[2]{\left|\left |#1 \right| \right |_{#2}}

\newtheorem{Theorem}{Theorem}

\newtheorem*{conj*}{Conjecture}
\newtheorem{Lemma}[Theorem]{Lemma}

\graphicspath{ {./images/} }

\title{(In)stability of de Sitter Quasinormal Mode spectra}
\author{C.~M.~Warnick}
\address{Centre for Mathematical Sciences, Wilberforce Road, Cambridge, CB3 0WA}
\email{c.m.warnick@maths.cam.ac.uk}
\date{\today}

\begin{document}
\begin{abstract}
    We consider how the quasinormal spectrum for the conformal wave operator on the static patch of de Sitter changes in response to the addition of a small potential. Since the quasinormal modes and co-modes are explicitly known, we are able to give explicit formulae for the instantaneous rate of change of each frequency in terms of the perturbing potential. We verify these exact computations numerically using a novel technique extending the spectral hyperboloidal approach of Jaramillo et al.\ (2021). We propose a definition for a family of pseudospectra that we show capture the instability properties of the quasinormal frequencies.
\end{abstract}

\maketitle

\section{Introduction}
For asymptotically de Sitter and anti-de Sitter black hole spacetimes, the problem of defining the quasinormal frequencies has been satisfactorily resolved based on making use of a hyperboloidal foliation of the spacetime \cite{Vasy2013MicrolocalDyatlov, Warnick2015OnHoles}\footnote{see \cite{Dyatlov2019MathematicalResonances, Gajic2021QuasinormalSpacetimes} and references therein for a historical overview of this problem}. For asymptotically flat black hole spacetimes the situation is not as fully developed, but nevertheless in many cases a suitably robust mathematical definition exists either through casting the problem in terms of scattering resonances and making use of the method of complex scaling \cite{Dyatlov2019MathematicalResonances, Stucker2024QuasinormalHole}, or through using a hyperboloidal slicing \cite{Gajic2021QuasinormalSpacetimes, Gajic2020AHoles, Gajic2024QuasinormalSpacetimes}. In all cases, the quasinormal frequencies can ultimately be understood as eigenvalues of some operator which is not self-adjoint.

A feature of operators which are not self-adjoint is that their spectra can be unstable to `small' perturbations. For a simple example in finite dimensions, consider the matrices
\[
A = \begin{pmatrix}
0 & \epsilon^{-3} \\
0 & 0
\end{pmatrix}, \qquad A' = \begin{pmatrix}
0 & \epsilon^{-3} \\
\epsilon & 0
\end{pmatrix}.
\]
Clearly for $0<\epsilon \ll 1$, $A' - A$ is `small' by any reasonable notion of smallness, however $A$ has a repeated eigenvalue at $0$, while $A'$ has eigenvalues $\pm \epsilon^{-1}$, so the spectra diverge as $\epsilon \to 0$.

In the context of black hole quasinormal spectra, it was noticed already by Aguirregabiria--Vishveshwara Nollert--Price and  in the `90s \cite{Aguirregabiria1996ScatteringApproach, Vishveshwara1996OnJourney,Nollert1996AboutHoles, Nollert1999QuantifyingSystems} that seemingly innocuous changes to the operators used in defining quasinormal modes could have dramatic effects on the spectrum. Motivated in part by mathematical \cite{Warnick2015OnHoles, Gajic2021QuasinormalSpacetimes,Gajic2020AHoles, Bizon2020AModes} and numerical \cite{Ansorg2016SpectralSlices,PanossoMacedo2018HyperboloidalCase,PanossoMacedo2020HyperboloidalSpacetime} works which cast the problem of finding the quasinormal spectrum as an eigenvalue problem for the time evolution operator on a hyperboloidal foliation, there has been a resurgence of interest in the problem of quasinormal spectral stability, see \cite{Jaramillo2021PseudospectrumInstability, AlSheikh2022ScatteringSystems} in the specific context of instability arising from non-self adjoint operators as well as \cite{Cheung2022DestabilizingFlea,Sarkar2023PerturbingConstant,Arean2023PseudospectraModes, Destounis2023Black-holePseudospectrum, Cownden2024TheAdS, Boyanov2024StructuralPseudospectrum} and references therein for many other works.

In this short article we shall consider the problem of the conformal wave equation on the static patch of de Sitter space. The high degree of symmetry enjoyed by the de Sitter spacetime means that the problem of determining the quasinormal spectrum is completely solvable, and the various objects involved can be computed explicitly. This makes this a helpful test-bed for understanding the effects on the spectrum of small perturbations. The perturbations we consider consist of stationary modifications to the potential. In a more physically motivated situation, we should consider the linearised gravitational field (rather than a conformal scalar field) and permit perturbations to the geometry of the background rather than just a potential. Our methods can, in principle, be applied in this situation but for simplicity we focus on the toy model.\footnote{A further complication can arise where the perturbations are presumed to have a time dependence with a typical timescale much shorter than the quasinormal frequencies, in which case one may hope to attempt some averaging procedure (see \cite[\S 3.5 d)]{Jaramillo2022PseudospectrumTransientsb}) }

We are able to compute exactly the first order correction to each quasinormal frequency in terms of the perturbing potential. We find that the answer to the question of whether an individual quasinormal frequency is stable to `small' perturbations depends sensitively on what is meant by `small' (cf \cite[footnote 18]{Gasperin2022EnergyProduct}). In particular, the relevant notion of smallness varies depending on which modes we are considering, and those representing more rapidly decaying modes require a more stringent notion of smallness. One may alternatively view this by first fixing the notion of smallness considered and then one observes that the more rapidly decaying modes are more unstable to small perturbations, consistent with expectations going back to \cite{Vishveshwara1996OnJourney}\footnote{\emph{``...we find that the fundamental mode is, in general, insensitive to small changes in the potential, whereas the higher modes could alter drastically.''}}.

In order to confirm the analytic computations, we also perform some numerics. For this we make use of a spectral method on a hyperboloidal (or null) slicing, similar to that used in \cite{Jaramillo2021PseudospectrumInstability}, but applied to an enlarged system obtained by differentiating the equation by hand $k$-times motivated by the analysis of \cite{Warnick2015OnHoles}. This has a doubly beneficial effect -- firstly it stabilises the numerical computation of quasinormal frequencies; secondly it permits us to stably compute a family of pseudospectra that we define, associated with the problem, which allow the stability properties to be directly visualised. In this context we should also mention the forthcoming work \cite{BessonBlackPreparation} which also provides a numerically stable computation of pseudospectra.

\section{Set-up and defining the quasinormal spectrum}
We consider the static patch of the de Sitter spacetime, written in coordinates that are regular at the future horizon. This is a metric on $\R^4 = \{(t, \bm{x}):t \in \R, \bm{x}\in \R^3 \}$
\begin{equation}
    g = -\left( 1-\kappa^2 \delta_{ij} x^ix^j \right) dt^2 - 2 \kappa \delta_{ij} x^i dx^j  dt + \delta_{ij}dx^i dx^j
\end{equation}
with $\delta_{ij}$ the usual Kronecker delta and $\kappa >0$ a constant. The static patch is the region $\mathcal{R} = \R_t \times B$ where $B = \{\bm{x} \in \R^3: |\bm{x}|<{\kappa^{-1}}\}$ is the ball of radius $\kappa^{-1}$, and the future cosmological horizon is $\mathcal{H}_+ = \R_t \times \partial B$. This metric is Einstein with cosmological constant $\Lambda = 3\kappa^2$. We will keep track of $\kappa$ for later discussion, but nothing is lost by setting $\kappa = 1$ throughout.

The wave operator in these coordinates takes the form
\[
\Box_g \psi = -\left(\frac{\partial}{\partial t} + \kappa x^i \frac{\partial}{\partial x^i} \right)^2 \psi  - 3 \kappa \left(\frac{\partial}{\partial t} + \kappa x^i \frac{\partial}{\partial x^i} \right) \psi +\delta^{ij} \frac{\partial^2 \psi } {\partial x^i \partial x^j}.
\]
We shall consider the following family of equations on this background
\begin{equation}\label{LDef}
    L(h)\psi := \Box_g \psi  - \kappa^2 V_h \psi= 0.
\end{equation}
Here $V_h$ is a time-independent potential depending on some small parameter $|h|<\epsilon$, and we assume that the map $(h, \bm{x}) \mapsto V_h(\bm{x})$ is smooth on $(-\epsilon, \epsilon)\times \R^3$. 
We are interested in particular in the quasinormal ring-down behaviour of solutions to this equation. To discuss this, we introduce the Laplace transformed operator which acts on functions $u:B \to \C$
\begin{equation}
    \hat{L}(s, h) u := e^{-s t} L(h)(e^{s t} u).
\end{equation}

We define the quasinormal frequencies through the solvability properties of this operator. More precisely, for $k = 0, 1, 2, \ldots$ we define an inner product and norm on functions $u, w :B \to \C$ by
\begin{equation}
    (u, w)_k := \sum_{l=0}^k \int_{B} \left(\overline{\nabla^{(l)} u} \cdot \nabla^{(l)} w \right)\, d^3x, \qquad\qquad \norm{u}{k} := (u, u)_k^{\frac{1}{2}}.
\end{equation}
Here $\nabla^{(l)} u$ is the rank $l$-tensor $\nabla_{i_1}\cdots \nabla_{i_l}u$, and $\cdot$ means contraction on all indices\footnote{derivatives should be understood in the distributional sense}. Notice that $(u,w)_0$ is the usual $L^2-$inner product. We define $H^k$, the Sobolev space of order $k$, to consist of those functions $u:B\to \C$ with $\norm{u}{k}<\infty$. This is a Hilbert space with the corresponding inner product. We define the domain of $L_s$ to be
\[
D^k = \{u \in H^{k}: \hat{L}(1,0) u \in H^{k}\}.
\]
It can be shown that $H^{k+2}\subset D^k \subset H^{k+1}$, so that $u \in D^k \implies \hat{L}(s,h) u \in H^{k}$ for all $s, h$.

With this definition in hand, we can state the basic theorem we shall require, which follows straightforwardly from \cite{Vasy2013MicrolocalDyatlov, Warnick2015OnHoles, Hintz2021QuasinormalSpace, Hintz2022QuasinormalHoles}:
\begin{Theorem} \label{Fred1}
    Let $U_k = \{z\in \C: \Re(z) >-(k+\frac{1}{2})\kappa\}$, and fix $h$. Then the operator $\hat{L}(s,h) : D^k \to H^{k}$ is invertible for $s \in U_k$, except at a discrete set $\Lambda_k(h) \subset U_k$. Moreover, for each $\sigma \in \Lambda_k(h)$ there is an integer $d>0$ such that:
    \begin{enumerate}[i)]
        \item There exists a $d$-dimensional space of smooth functions $w:B\to \C$ which extend smoothly to $\partial B$ and satisfy $\hat{L}(\sigma,h) w = 0$.
        \item There exists a $d$-dimensional space of distributions $X \in \mathscr{D}'(\R^3)$ which satisfy
        \begin{equation}\label{ComodeDef}
        X[\hat{L}(\sigma,h) \phi] = 0, \qquad \qquad  |X[\phi]| \leqslant c\norm{\phi|_{B}}{k},
        \end{equation}
        for some $c>0$ and all test functions $\phi \in C^\infty_c(\R^3)$.
        \item As $s$ varies, the meromorphic family of operators $\hat{L}(s,h)^{-1} : H^{k} \to D^k$ has a pole at $\sigma$.
    \end{enumerate}
\end{Theorem}
It follows from the characterisation of points in $\Lambda_k(h)$ that $\Lambda_{k+1}(h)\cap U_k = \Lambda_{k}(h)$. We call any $\sigma\in \Lambda_k(h)$ for some $k$ a quasinormal frequency of $L(h)$, with geometric multiplicity $d$. A corresponding smooth solution to $\hat{L}(\sigma,h) w = 0$ is a quasinormal mode, and a distribution $X$ satisfying $ii)$ above we call a co-mode. Notice that the condition on $X$ implies that $X$ is supported in $\overline{B}$, and so can be uniquely extended to act on test functions in $C^\infty(\overline{B})$. 

The residue of $\hat{L}(s,h)^{-1}$ at $s=\sigma$ is a finite rank operator, and we identify the rank of this residue with the algebraic multiplicity of $\sigma$. As in the familiar case of matrices, the algebraic multiplicity is an upper bound for the geometric multiplicity. We say that a quasinormal frequency $\sigma \in \Lambda_k(h)$ is \emph{simple} if it has algebraic multiplicity one.

The result above holds for $h$ fixed. The question we shall consider in this paper, that of quasinormal spectral instability, amounts to trying to understand how the set $\Lambda_k(h)$ changes as $h$ varies. 

\section{Stability of quasinormal frequencies}

\subsection{Simple quasinormal frequencies}

Let us suppose that for the unperturbed operator, i.e.\ at $h=0$, we can compute the quasinormal frequencies, modes and co-modes, and we consider some simple $\sigma \in \Lambda_k(0)$ with corresponding quasinormal mode $w$ and co-mode $X$. It was shown in \cite{Joykutty2022ExistenceHoles} that that as $h$ varies there is some smooth curve of quasinormal frequencies $\sigma(h) \in \Lambda_k(h)$, with $\sigma(0) = \sigma$, together with an associated curve of quasinormal modes $w(h)$ with $w(0) = w$, depending smoothly on $h$ such that
\[
\hat{L}( \sigma(h),h) w(h) = 0
\]
holds for all $|h|<\epsilon$. Moreover, in \cite{Joykutty2022ExistenceHoles} an explicit power series expansion for $\sigma(h)$ is given in terms of the trace of certain operator valued contour integrals. We shall take a more elementary approach to find a formula for $\sigma'(0)$.

Since $\hat{L}$ depends smoothly on its arguments, we can differentiate with respect to $h$ at $h=0$ to find:
\begin{equation} \label{diffcond}
    \sigma'(0) \frac{\partial \hat{L}}{\partial s}(\sigma, 0) w + \frac{\partial \hat{L}}{\partial h}(\sigma, 0) w + \hat{L}(\sigma, 0) w'(0) = 0.
\end{equation}
By assumption, we know $\frac{\partial \hat{L}}{\partial s}(\sigma, 0), \frac{\partial \hat{L}}{\partial h}(\sigma, 0)$ and $w$, but we don't know anything about $w'(0)$. If, however, we act on \eqref{diffcond} with the co-mode $X$, the term involving $w'(0)$ will be annihilated. We find then:
\[
\sigma'(0) X\left[\frac{\partial \hat{L}}{\partial s}(\sigma, 0) w \right] + X\left[\frac{\partial \hat{L}}{\partial h}(\sigma, 0) w\right] = 0
\]
or, rearranging
\begin{equation}\label{Sigdot}
\sigma'(0)  = -\frac{X\left[\frac{\partial \hat{L}}{\partial h}(\sigma, 0) w\right]}{X\left[\frac{\partial \hat{L}}{\partial s}(\sigma, 0) w \right]}.
\end{equation}
This formula gives us an exact expression for the velocity of the curve of quasinormal frequencies $\sigma(h)$ as it passes through $\sigma$. 

We observe that only the numerator of \eqref{Sigdot} depends on the perturbation -- the denominator can be computed from the unperturbed operator alone. Recalling that the operator norm of a linear map $A : V \to W$ between normed spaces is given by
\[
\|A\|_{V \to W} = \sup_{u \in V, \|u\|_V = 1} \|Au\|_{W},
\]
we can estimate $\sigma'(0)$ in terms of an operator norm of the linearised perturbation as
\begin{equation}\label{Sigdotest}
|\sigma'(0)| \leqslant \gamma_\sigma \left \| \frac{\partial \hat{L}}{\partial h}(\sigma, 0) \right \|_{H^k \to H^k}.
\end{equation}
Here the sensitivity, or condition number, $\gamma_\sigma$ depends only on the unperturbed operator, and is given by
\[
\gamma_\sigma = \frac{\|w\|_{k} \|X\|_{k*}}{\left|X\left[\frac{\partial \hat{L}}{\partial s}(\sigma, 0) w \right] \right|}
\]
where $\|X\|_{k*} := \|X\|_{H^k \to \C}$. We can think of the expression for $\gamma$ as a generalisation of the formula for the sensitivity of a matrix eigenvalue. 

At this stage, it is worth commenting on the role of $k$ in the discussion. Increasing $k$ increases the region of the complex plane in which we can study the quasinormal frequencies, however the price we pay for this in \eqref{Sigdotest} is an increase in the control that we require on the perturbation. We can mitigate this by choosing $k$ to be as small as possible, consistent with $\sigma \in \Lambda_k(0)$. Even doing this we see that to bound the rate of change of a quasinormal frequency $\sigma$ we (roughly speaking) need control of more than $-\kappa^{-1} (\textrm{Re }\sigma)$ derivatives of the perturbation. We shall see this more explicitly later on. We should note that `control of higher derivatives' may appear to be an unphysical condition, but one can also view this condition as asking that the perturbations should not have too much of their energy at high wavenumbers\footnote{Roughly speaking, for a perturbation in $H^k$, the fraction of the total energy carried by wavenumbers greater than $\mu$ is bounded by a constant multiple of $\mu^{-2k}$ for large $\mu$.}. 

The arguments above do not rely strongly on the particular form of the metric, or the family of operators we consider. As long as a result broadly analogous to the conclusions of Theorem \ref{Fred1} holds, we can expect to be able to repeat this argument.

\subsection{Non-simple quasinormal frequencies}

In the discussion above, we made the assumption that the quasinormal frequency $\sigma$ was simple, which was needed in order to establish that $\sigma$ sits on a smooth curve of quasinormal frequencies. If $\sigma$ is not simple, then this need not be the case in general - see Figure \ref{Plot3} for a situation where this arises in our numerics. It does follow from \cite{Joykutty2022ExistenceHoles} that the number of QNFs, counted with suitable multiplicity, inside a small circle around $\sigma$ is independent of $h$ for small $h$, so that QNFs in particular cannot be locally `created' or `destroyed' by small perturbations of the type we consider -- QNFs can only appear from infinity or by splitting off from a QNF with algebraic multiplicity greater than one.

In general it does not appear to be a straightforward question to determine whether a particular non-simple quasinormal frequency lies on a smooth curve. In some cases, however, it may be that evolution under $L(h)$ leaves invariant some subspace (such as an angular momentum eigenspace) so we can consider the problem of finding quasinormal frequencies restricted to this subspace. If $\sigma$ is a simple quasinormal frequency of the restricted problem, then the results of the previous section will apply.

\section{The generalised pseudo-spectrum}

In \cite{Jaramillo2021PseudospectrumInstability} and subsequently (see \cite{ Boyanov2024StructuralPseudospectrum} and references therein), the instability of the quasinormal spectrum has been investigated using the notion of pseudospectrum, comparing results from this approach to computations with explicit perturbations. Recall that for a matrix $A$ we can define the $\epsilon-$pseudospectrum to be\footnote{The pseudospectrum is usually defined as a closed set, with $\geqslant$ in place of $>$, however the open condition generalises more straightforwardly to the infinite dimensional case.}
\[
\Lambda^\epsilon = \{ s \in \C : \norm{(A-s \iota)^{-1}}{\R^n \to \R^n} > \epsilon^{-1}\}.
\]
where we define $\norm{(A-s \iota)^{-1}}{\R^n \to \R^n}= \infty$ whenever $A-s \iota$ is not invertible. It can be shown \cite{Trefethen1999SpectraPseudospectra,EmbreePseudospectraGateway, vanDorsselaer1993LinearProblems} that $\Lambda^\epsilon$ corresponds to the set of points which can appear in the spectrum of $A+ \delta A$, where $\delta A$ is a perturbation satisfying $\norm{\delta A}{\R^n \to \R^n}<\epsilon$.

This notion generalises to operators on infinite dimensional spaces in the obvious way. However, this definition cannot immediately be applied to our problem above because $\hat{L}(s, h)$ is not of the form $A - s I$ for some operator $A$. There are two possible approaches to resolve this. The approach taken by \cite{Jaramillo2021PseudospectrumInstability} is to follow \cite{Warnick2015OnHoles, Ansorg2016SpectralSlices, PanossoMacedo2018HyperboloidalCase} and recast the problem of finding the quasinormal frequencies as a genuine eigenvalue problem by writing
\[
\hat{L}(s, h) = L_2(h) + s L_1(h) + s^2
\]
where $L_j(h)$ is a differential operator of order $j$. Then we can verify that $\hat{L}(s, h)w = 0$ has a solution if and only if
\[
\begin{pmatrix}
-s & 1\\
-L_2(h) & -L_1(h)-s
\end{pmatrix}\begin{pmatrix}
w \\
v 
\end{pmatrix} =0 
\]
has a smooth solution. Thus the set $\Lambda_k(h)$ can be identified with the part of the spectrum of
\[
\mathcal{L}(h) = \begin{pmatrix}
0 & 1\\
-L_2(h) & -L_1(h)
\end{pmatrix}
\]
in $U_k$, where $\mathcal{L}(h)$ is thought of as a closed unbounded operator on $\mathcal{H}^k:=H^{k+1} \times H^{k}$. This motivates one definition of the $\epsilon-$pseudospectrum\footnote{The pseudospectrum is a property of the \emph{unperturbed} operator, hence we set $h=0$.} as
\[
\tilde{\Lambda}_k^\epsilon = \{ s \in \C : \norm{(\mathcal{L}(0)-s \iota)^{-1}}{\mathcal{H}^k \to \mathcal{H}^k} > \epsilon^{-1}\}.
\]
This has the advantage of being the standard definition applied to $\mathcal{L}$, but the disadvantage that in numerical computations one has to double the dimension of the approximation space to account for the two functions $w, v$. Moreover, since $\mathcal{L}$ does not have compact resolvent, approximation by matrices can be more challenging.

An alternative approach is to generalise the notion of $\epsilon-$pseudospectrum by declaring
\begin{equation} \label{PSdef}
    {\Lambda}_k^\epsilon = \{ s \in U_k : \norm{\hat{L}(s,0)^{-1}}{H^{k} \to H^{k}} > \epsilon^{-1}\}.
\end{equation}
This has the advantage that $\hat{L}(s,0)^{-1}:H^{k} \to H^{k}$ is compact, but the disadvantage that since it is not the standard definition of pseudospectrum one cannot readily make use of existing numerical libraries. We note as an aside that we could also consider the $H^{k} \to D^k$ norm in place of the $H^{k} \to H^{k}$ norm in \eqref{PSdef}, but it will not make a significant difference for the type of perturbations we consider.

We shall take \eqref{PSdef} as our definition of the pseudospectrum for the rest of the paper (see \cite{BessonBlackPreparation} for an alternative approach). A modification of the usual arguments for pseudospectra \cite{EmbreePseudospectraGateway,vanDorsselaer1993LinearProblems} shows that ${\Lambda}_k^\epsilon$ is precisely the set of points in $U_k$ which can occur as quasinormal frequencies of $L(s, 0) + E$ for some operator $E:H^k \to H^k$ satisfying $\norm{E}{H^{k} \to H^{k}}<\epsilon$. One can verify that the fact that $s$ does not appear linearly in $L(s, 0)$ does not affect this argument. In particular, provided $\norm{V_h}{C^k}<\epsilon/\kappa^2$ we have $\Lambda_k(h) \subset {\Lambda}_k^\epsilon$.

We note that this definition agrees with that for the null-slicing in \cite{Cownden2024TheAdS, Boyanov2024StructuralPseudospectrum}, however we don't assume that the slicing is everywhere null.

\section{Explicit computations for perturbations of the conformal wave operator}\label{CompSec}

In order to give a concrete demonstration of the ideas above, we will work in a setting where the quasinormal frequencies, modes and co-modes of the operator are known explicitly at $h=0$. In particular, from now on we assume that we perturb about the conformal wave operator on de Sitter, in our language:
\[
V_0(\bm{x}) = 2.
\]
Under this assumption, we have \cite{Hintz2021QuasinormalSpace, Hintz2022QuasinormalHoles}:
\begin{Lemma}
    Suppose $V_0(\bm{x}) = 2$. Then:
    \begin{enumerate}[i)]
    \item $\Lambda_{k}(0) = \{-\kappa, -2 \kappa, -3\kappa,\ldots,  -k\kappa \}$.
    \item The quasinormal frequency $\sigma_n := -n \kappa \in \Lambda_{k}(0)$ has geometric and algebraic multiplicity $n^2$, and a basis for the corresponding space of quasinormal frequencies is given in terms of the standard spherical polar coordinates $(r, \theta, \phi)$ on $B_\kappa$ by
    \[
    w_{n, l, m} = (\kappa r)^l Y_{l,m}(\theta, \phi) {}_2F_1\left[\frac{1+l-n}{2}, \frac{2+l-n}{2}, \frac{3+2l}{2};  \kappa^2 r^2 \right].
    \]
    Here $Y_{l, m}$ are the spherical harmonics, ${}_2F_1$ is the hypergeometric function and the integers $m, l$ satisfy $|m|\leqslant l\leqslant n$.
    \item For each $\sigma_n \in \Lambda_{k}(0)$, the corresponding quasinormal co-modes are supported on $\partial B$. A basis for the space of co-modes is given in terms of the action on a smooth test function by
    \begin{equation}\label{XnlmDef}
    X_{n, l, m}[\varphi] = \sum_{i=0}^{n-1} A^i_{n, l}    \frac{1}{\kappa^i} \left. \frac{d^i \varphi_{l,m}}{d r^i}\right|_{r=\kappa^{-1}}
    \end{equation}
    where $|m|\leqslant l\leqslant n$, $A^i_{n, l}$ are constants, and $\varphi_{l,m}(r)$ is the projection of $\varphi$ onto the $(l, m)-$spherical mode.
    \end{enumerate}
\end{Lemma}

While it is possible to specify the constants $A^i_{n, l}$ explicitly, see \cite{Hintz2021QuasinormalSpace, Hintz2022QuasinormalHoles, Joykutty2024QuasinormalHoles}, for the purposes of our results below it is more computationally efficient to find $X_{n, l, m}$ for any particular choice of $n, l, m$ by simply using \eqref{XnlmDef} as an ansatz in \eqref{ComodeDef} and solving the resulting linear system for $A^i_{n,l}$. Doing so using Mathematica to perform the computations, we find the results in Table \ref{tab:Ainl}.

\begin{table}[h]
    \begin{subtable}[h]{0.45\textwidth}
        \centering
\begin{NiceTabular}{|cc|cccccc|}
\Hline
\Block{2-2}{\diagbox{$\,n$}{$i\,$}} &&
\Block{2-1}{0}   & 
\Block{2-1}{1}   & 
\Block{2-1}{2}   & 
\Block{2-1}{3} &
\Block{2-1}{4} &
\Block{2-1}{5} \\
\\
\Hline 
\Block{1-2}{1}  &&1&&&&\\
\Block{1-2}{2}  &&1&1&&&\\
\Block{1-2}{3}  &&0&2&1&&\\
\Block{1-2}{4}  &&0&0&3&1&\\
\Block{1-2}{5}  &&0&0&0&4&1\\
\Block{1-2}{6}  &&0&0&0&0&5&1\\
\Hline
\end{NiceTabular}
\caption{$l=0$}
  \end{subtable}
    \begin{subtable}[h]{0.45\textwidth}
    \centering
\begin{NiceTabular}{|cc|ccccccc|}
\Hline
\Block{2-2}{\diagbox{$\,n$}{$i\,$}} &&
\Block{2-1}{0}   & 
\Block{2-1}{1}   & 
\Block{2-1}{2}   & 
\Block{2-1}{3} &
\Block{2-1}{4} &
\Block{2-1}{5} &
\Block{2-1}{6} \\
\\
\Hline 
\Block{1-2}{2}  &&2&1&&&\\
\Block{1-2}{3}  &&0&3&1&&\\
\Block{1-2}{4}  &&0&0&4&1&\\
\Block{1-2}{5}  &&0&0&0&5&1\\
\Block{1-2}{6}  &&0&0&0&0&6&1\\
\Block{1-2}{7}  &&0&0&0&0&0&7&1\\
\Hline
\end{NiceTabular}
\caption{$l=1$}
 \end{subtable}
 
\vspace{0.5cm}
\begin{subtable}[h]{\textwidth}
    \centering
\begin{NiceTabular}{|cc|cccccccc|}[columns-width=auto]
\Hline
\Block{2-2}{\diagbox{\quad$\,n$}{$i\,$\quad${}$}} &&
\Block{2-1}{0}   & 
\Block{2-1}{1}   & 
\Block{2-1}{2}   & 
\Block{2-1}{3} &
\Block{2-1}{4} &
\Block{2-1}{5} &
\Block{2-1}{6} &
\Block{2-1}{7}\\
\\
\Hline 
\Block{1-2}{3}  &&3&5&1&\\
\Block{1-2}{4}  &&$-3$&3&6&1\\
\Block{1-2}{5}  &&6&$-6$&3&7&1\\
\Block{1-2}{6}  &&$-18$&18&$-9$&3&8&1\\
\Block{1-2}{7}  &&72&$-72$&36&$-12$&3&9&1\\
\Block{1-2}{8}  &&$-360$&360&$-180$&60&$-15$&3&10&1\\
\Hline
\end{NiceTabular}
\caption{$l=2$}
     \end{subtable}
     \caption{The coefficients $A^i_{n,l}$ for the first six quasinormal frequencies with in each of the angular sectors $l=0, 1, 2$.}
     \label{tab:Ainl}
\end{table}

Since for $\sigma \neq -\kappa$ the quasinormal frequencies are not simple, in order to make use of \eqref{Sigdot} to estimate the change in the QNF we shall make the additional assumption that the potential $V_h$ is spherically symmetric. Under this assumption, the QNFs are simple once we restrict our attention to a single fixed angular mode. If we fix $l,m$ with $|m|\leqslant l$, then for $k\geqslant l$, the unperturbed quasinormal spectrum restricted to the $l, m$ angular mode is $\Lambda_k^{l,m}(0) = \{-l \kappa, \ldots, -k \kappa\}$ and all the QNFs are simple. We can compute the rate of change of the QNF at $-\kappa n$ by
\[
\sigma'_{n, l, m}(0)  = -\frac{X_{n, l, m}\left[\frac{\partial \hat{L}}{\partial h}(\sigma, 0) w_{n, l, m}\right]}{X_{n, l, m}\left[\frac{\partial \hat{L}}{\partial s}(\sigma, 0) w_{n, l, m} \right]}.
\]

In order to use this formula we also need $\frac{\partial \hat{L}}{\partial s}$ and $\frac{\partial \hat{L}}{\partial h}$. For the particular case of interest, with $L(h)$ given by \eqref{LDef}, we have
\[
\hat{L}(s,h) u = -\left(s +  \kappa x^i \frac{\partial}{\partial x^i} \right)^2 u  - 3 \kappa \left(s +  \kappa x^i \frac{\partial}{\partial x^i} \right) u +\delta^{ij} \frac{\partial^2 u } {\partial x^i \partial x^j} - \kappa^2 V_h u
\]
so that
\[
\frac{\partial \hat{L}}{\partial s}(\sigma, 0) u = -2 \kappa x^i \frac{\partial u}{\partial x^i} - (2s + 3 \kappa) u =-2 \kappa r \frac{\partial u}{\partial r} - (2s + 3 \kappa) u .
\]
and
\[
\frac{\partial \hat{L}}{\partial h}(\sigma, 0) u = - \kappa^2 W u,
\]
where we introduce $W = \left.\frac{\partial V_h}{\partial h} \right|_{h=0}$, the first order perturbation to the potential. 

We now have all that's required to compute $\sigma'_{n, l, m}(0)$. In view of the structure of the operator $X_{n, l, m}$, we can write
\[
\sigma'_{n, l, m}(0) = \kappa \sum_{i=0}^{n-1} B^i_{n, l}   \frac{1}{\kappa^i} W^{(i)}(\kappa^{-1})
\]
for some constants $B^i_{n, l}$. Note that this is independent of $m$ due to the spherical symmetry of the perturbing potential. We can again use Mathematica to compute these constants and present the results for the first few modes in the $l=0, 1, 2$ angular sectors in Table \ref{tab:Binl}.

\begin{table}[h!]
    \begin{subtable}[h]{\textwidth}
        \centering
\begin{NiceTabular}{|cc|rrrrrr|}[columns-width=auto, cell-space-limits=2pt]

\Hline
\Block{2-2}{\diagbox{\quad$\,n$}{$i\,$\quad${}$}} &&
\Block{2-1}{0}   & 
\Block{2-1}{1}   & 
\Block{2-1}{2}   & 
\Block{2-1}{3} &
\Block{2-1}{4} &
\Block{2-1}{5} \\
\\
\Hline 
\Block{1-2}{1} & & $-1$\\ 
\Block{1-2}{2} & & $1$ & $1$\\ 
\Block{1-2}{3} & & $-1$ & $-2$ & $-\frac{2}{3}$\\ 
\Block{1-2}{4} & & $1$ & $3$ & $2$ & $\frac{1}{3}$\\ 
\Block{1-2}{5} & & $-1$ & $-4$ & $-4$ & $-\frac{4}{3}$ & \
$-\frac{2}{15}$\\ 
\Block{1-2}{6} & & $1$ & $5$ & $\frac{20}{3}$ & $\frac{10}{3}$ & \
$\frac{2}{3}$ & $\frac{2}{45}$\\ 

\Hline
\end{NiceTabular}
\caption{$l=0$}
  \end{subtable}

\vspace{0.5cm}   
    \begin{subtable}[h]{\textwidth}
    \centering
\begin{NiceTabular}{|cc|rrrrrrr|}[columns-width=auto, cell-space-limits=2pt]
\Hline
\Block{2-2}{\diagbox{\quad$\,n$}{$i\,$\quad${}$}} &&
\Block{2-1}{0}   & 
\Block{2-1}{1}   & 
\Block{2-1}{2}   & 
\Block{2-1}{3} &
\Block{2-1}{4} &
\Block{2-1}{5} &
\Block{2-1}{6} \\
\\
\Hline 
\Block{1-2}{2} & & $-1$ & $-\frac{1}{3}$\\ 
\Block{1-2}{3} & & $1$ & $\frac{5}{3}$ & $\frac{1}{3}$\\ 
\Block{1-2}{4} & & $-1$ & $-\frac{41}{15}$ & $-\frac{8}{5}$ & \
$-\frac{1}{5}$\\ 
\Block{1-2}{5} & & $1$ & $\frac{19}{5}$ & $\frac{53}{15}$ & \
$\frac{16}{15}$ & $\frac{4}{45}$\\ 
\Block{1-2}{6} & & $-1$ & $-\frac{169}{35}$ & $-\frac{216}{35}$ & \
$-\frac{102}{35}$ & $-\frac{34}{63}$ & $-\frac{2}{63}$\\ 
\Block{1-2}{7} & & $1$ & $\frac{41}{7}$ & $\frac{199}{21}$ & \
$\frac{128}{21}$ & $\frac{110}{63}$ & $\frac{23}{105}$ & \
$\frac{1}{105}$\\ 
\Hline
\end{NiceTabular}
\caption{$l=1$}
 \end{subtable}
 
\vspace{0.5cm}
\begin{subtable}[h]{\textwidth}
    \centering
\begin{NiceTabular}{|cc|rrrrrrrr|}[columns-width=auto, cell-space-limits=3pt]
\Hline
\Block{2-2}{\diagbox{\quad$\,n$}{$i\,$\quad${}$}} &&
\Block{2-1}{0}   & 
\Block{2-1}{1}   & 
\Block{2-1}{2}   & 
\Block{2-1}{3} &
\Block{2-1}{4} &
\Block{2-1}{5} &
\Block{2-1}{6} &
\Block{2-1}{7}\\
\\
\Hline 
\Block{1-2}{3} & & $-1$ & $-\frac{3}{5}$ & $-\frac{1}{15}$\\ 
\Block{1-2}{4} & & $1$ & $\frac{11}{5}$ & $\frac{4}{5}$ & \
$\frac{1}{15}$\\ 
\Block{1-2}{5} & & $-1$ & $-\frac{117}{35}$ & $-\frac{93}{35}$ & \
$-\frac{64}{105}$ & $-\frac{4}{105}$\\ 
\Block{1-2}{6} & & $1$ & $\frac{157}{35}$ & $\frac{544}{105}$ & \
$\frac{227}{105}$ & $\frac{1}{3}$ & $\frac{1}{63}$\\ 
\Block{1-2}{7} & & $-1$ & $-\frac{583}{105}$ & $-\frac{887}{105}$ & \
$-\frac{316}{63}$ & $-\frac{82}{63}$ & $-\frac{1}{7}$ & \
$-\frac{1}{189}$\\ 
\Block{1-2}{8} & & $1$ & $\frac{139}{21}$ & $\frac{260}{21}$ & \
$\frac{601}{63}$ & $\frac{52}{15}$ & $\frac{194}{315}$ & \
$\frac{34}{675}$ & $\frac{1}{675}$\\ 
\Hline
\end{NiceTabular}
\caption{$l=2$}
     \end{subtable}
     \caption{The coefficients $B^i_{n,l}$ for the first six quasinormal frequencies within each of the angular sectors $l=0, 1, 2$.}
     \label{tab:Binl}
\end{table}

Picking two cases as examples, we can read off from the tables that:
\begin{align*}
\sigma_{1,0,0}'(0) &= - \kappa W(\kappa^{-1})\\
\sigma_{3, 1, 1}'(0) &= \kappa W(\kappa^{-1}) + \frac{5}{3} W'(\kappa^{-1})+ \frac{1}{3\kappa} W''(\kappa^{-1}).
\end{align*}
We see very explicitly here and from Table \ref{tab:Binl} that the rate of change of the quasinormal frequency depends on higher derivatives of the perturbing potential, and the larger $n$, i.e.\ the deeper into the stable plane we go, the more derivatives that are required. Equivalently: the deeper into the stable plane, the more control we require on the high wavenumber component of our perturbation. The 
increasing order of the operator norm that appears on the right-hand side of \eqref{Sigdotest} as we probe deeper into the plane is not simply an artefact of our framework, but is necessary.

To see why it is necessary to use higher order norms to constrain the perturbations, let us consider the case $\kappa = 1$ and consider a family of perturbations $W(r) = \epsilon^3 \exp(-\frac{r^2}{\epsilon^2})$. We clearly have that
\[
|W(r)| + |W'(r)| \lesssim \epsilon
\]
so in particular as $\epsilon \to 0$, we see that in the `energy norm', i.e.\ the operator norm associated to the $H^1$ norm we have that the perturbation tends to zero. However, $W''(1) \sim \epsilon^{-1}$ as $\epsilon\to 0$, so that (for example) the $l=m=1, n=3$ mode is displaced (to first order) by a term proportional to $\epsilon^{-1}$. Hence smallness of the perturbation in the energy norm is no guarantee of stability of the quasinormal modes lying sufficiently deep in the stable half-plane.

\begin{figure}
\centering
    \includegraphics[width=0.75\textwidth]{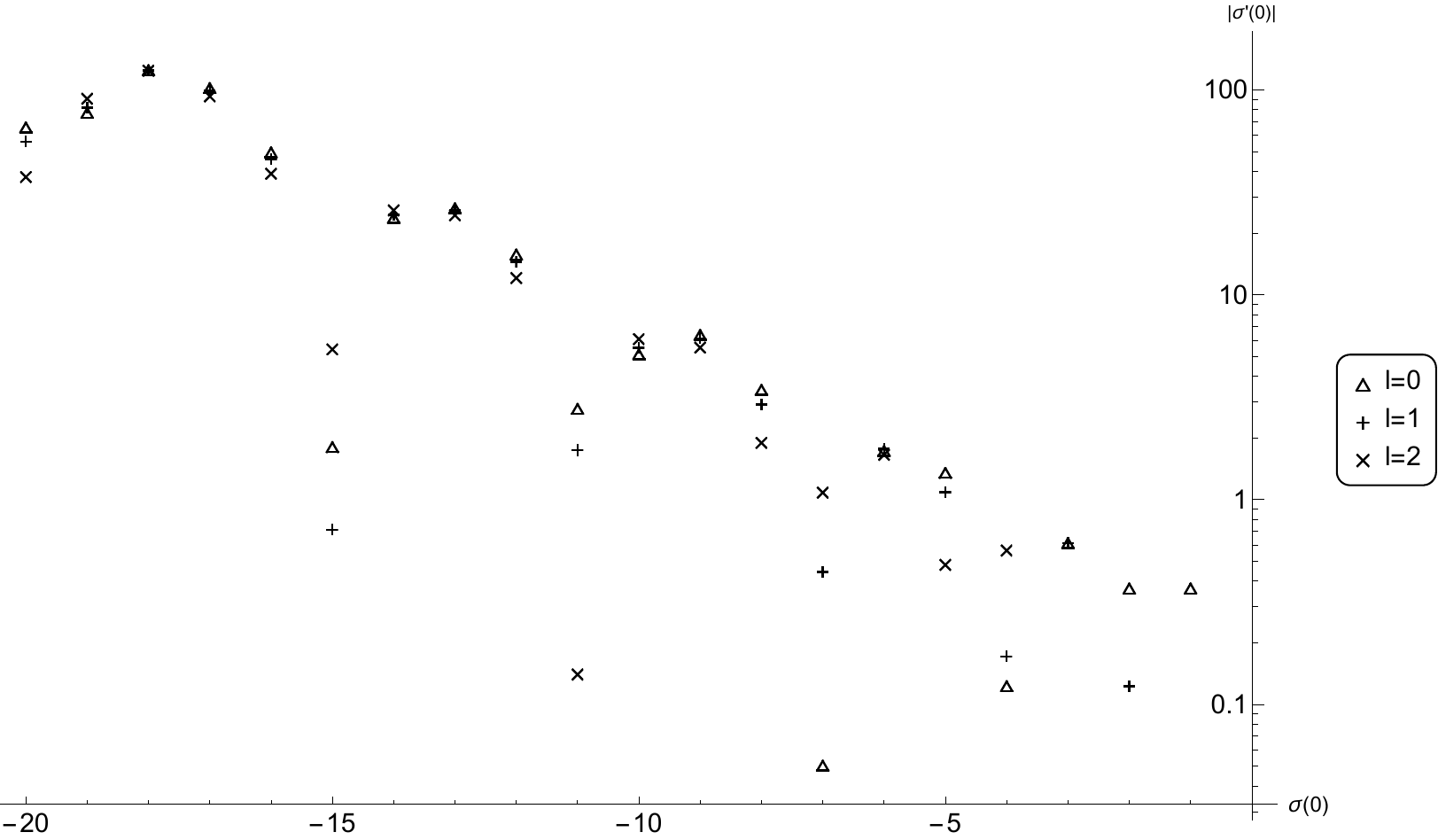}
    \caption{The magnitude of $\sigma'_{n, l, m}(0)$ for $n\leqslant 20$, $l\leqslant 2$ for the potential $V_h(r) = 2+h\exp(-r^2)$, with $\kappa = 1$.\label{Plot1}}
\end{figure}

For the choice of potential $V_h(r) = 2+h\exp(-r^2)$, with $\kappa = 1$, which we study numerically below, we have computed $\abs{\sigma'_{n, l, m}(0)}$ for $n\leqslant 20$, $l\leqslant 2$ and presented the results graphically. Noting the logarithmic scale on the $y-$axis, we see that for this choice of perturbing potential $\abs{\sigma'_{n, l, m}(0)}$ grows roughly exponentially in $n$, consistent with our expectation that modes deeper in the stable plane become more and more unstable.

\section{Numerical calculation of QNFs and comparison to analytic results}

In order to test numerically the computations above, we have computed the quasinormal frequencies for the choice $V_h(r) = 2+h\exp(-r^2)$. We first present the results, then comment on the methods used below.

\subsection{Results}
\begin{figure}
\begin{subfigure}{.5\textwidth}
  \centering
  \includegraphics[width=.9\linewidth]{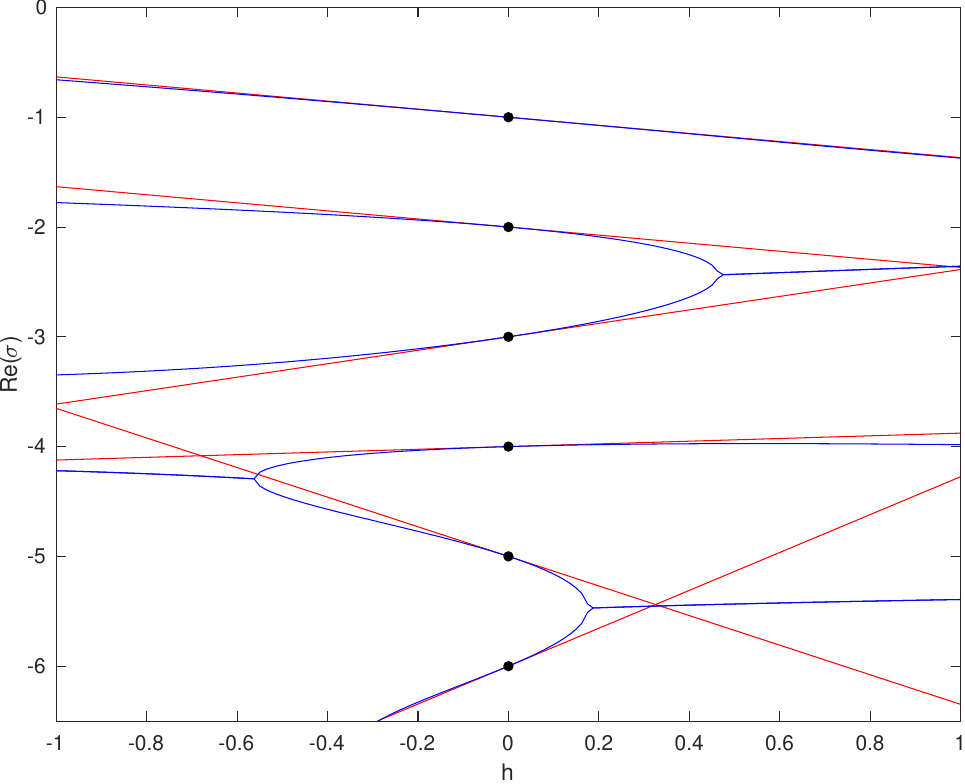}
  \caption{$l=0$}
  \label{Plot2:a}
\end{subfigure}%
\begin{subfigure}{.5\textwidth}
  \centering
  \includegraphics[width=.9\linewidth]{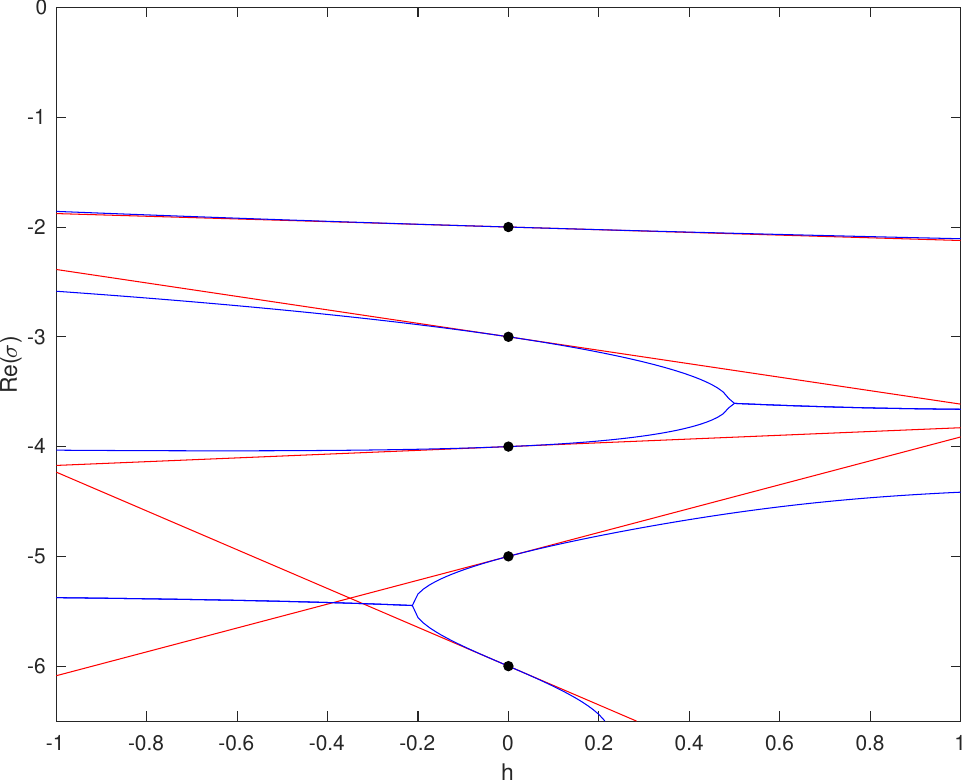}
  \caption{$l=1$}
  \label{Plot2:b}
\end{subfigure}
\begin{subfigure}{.5\textwidth}
  \centering
  \includegraphics[width=.9\linewidth]{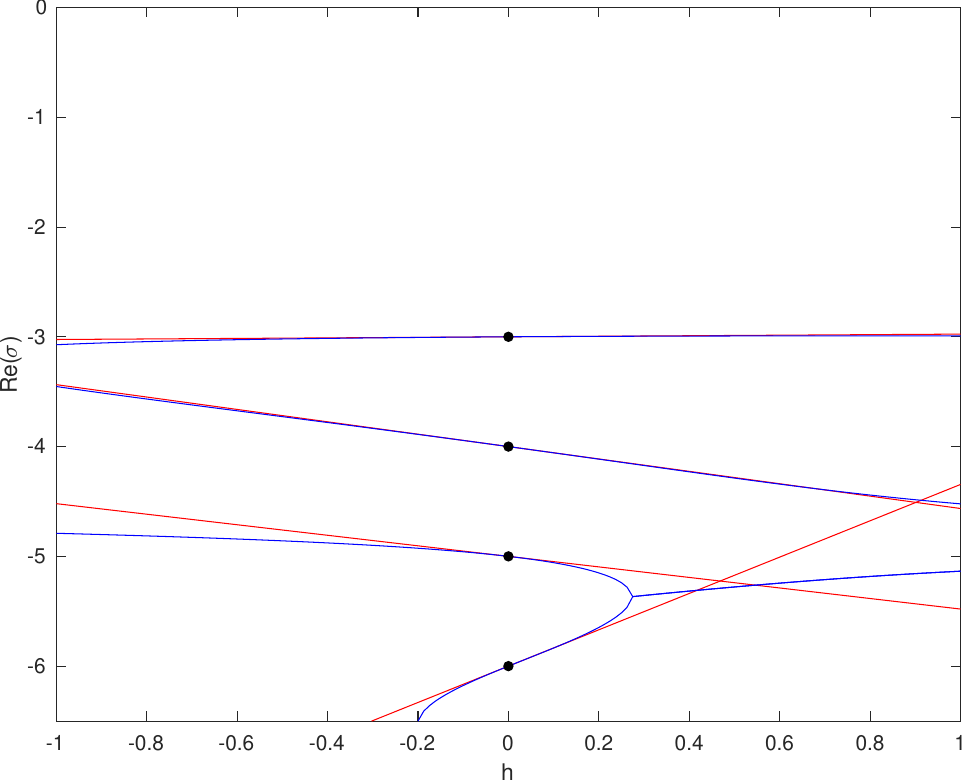}
  \caption{$l=2$}
  \label{Plot2:c}
\end{subfigure}%
\begin{subfigure}{.5\textwidth}
  \centering
  \includegraphics[width=.9\linewidth]{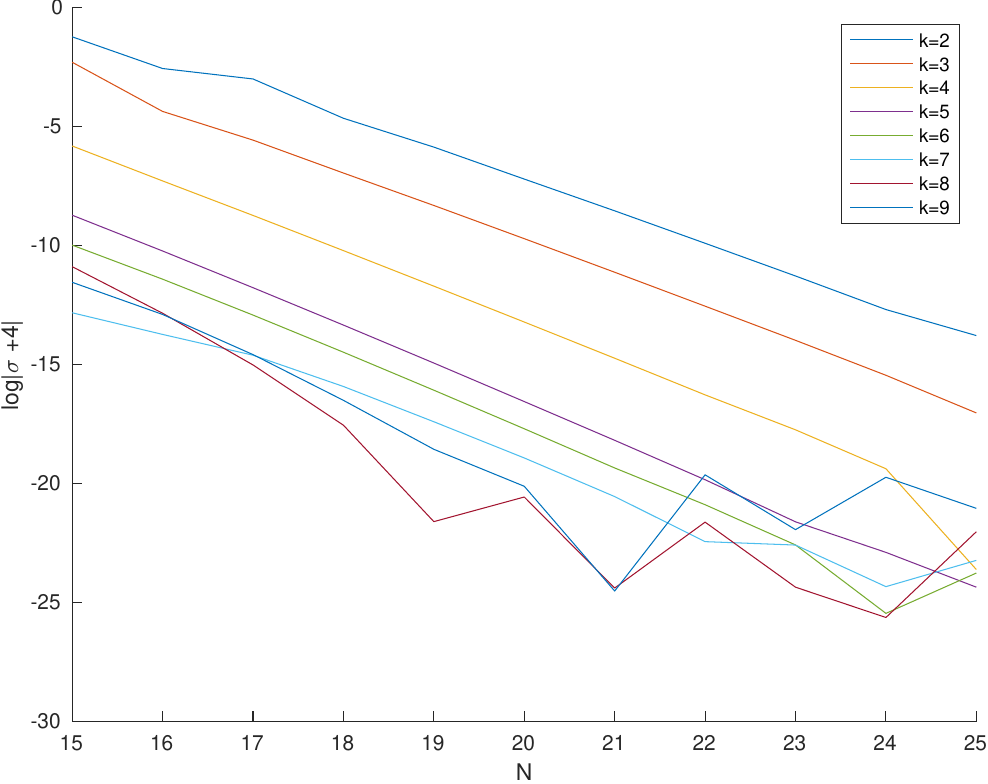}
  \caption{Logarithmic error against $N$ for $k=2, \ldots, 9$.}
  \label{Plot2:d}
\end{subfigure}
    \caption{$\textrm{Re}(\sigma(h))$ plotted against $h$ for numerically computed QNFs for $V_h = h \exp(-r^2)$ (blue lines) together with the linear approximations (red lines) in the $l=0, 1, 2$ sector. The black dots mark the location of the analytically known QNFs for $h=0$.\label{Plot2}}
\end{figure}

Since the equation is real, as is the quasinormal spectrum of $\hat{L}(s,0)$, frequencies can only move off the real axis in complex conjugate pairs. Restricted to each angular sector the QNFs are simple, so each QNF must remain real for a range of $h$ values near $0$. Accordingly, in Figure \ref{Plot2} we show the directly computed real part of the quasinormal frequencies as a function of $h$. Superposed on this we also plot the linear approximation to the QNFs given by
\[
\sigma_{n, l, m}(h) \approx \sigma_{n, l, m}(0)+h \sigma_{n, l, m}'(0),
\]
with $\sigma_{n, l, m}'(0)$ computed using the exact methods of \S\ref{CompSec} and we see very good agreement with the full numerical computation. We have experimented and this result is robust to changes to the potential, provided it remains smooth. We have thus verified the results of \S \ref{CompSec}. We note that this is a non-trivial test of our numerical scheme (described below) as it correctly predicts the values of the QNFs for $h=0$ and agrees with the analytical computations for the gradient of the blue curves at these points.

\begin{figure}
\centering
    \includegraphics[width=0.75\textwidth]{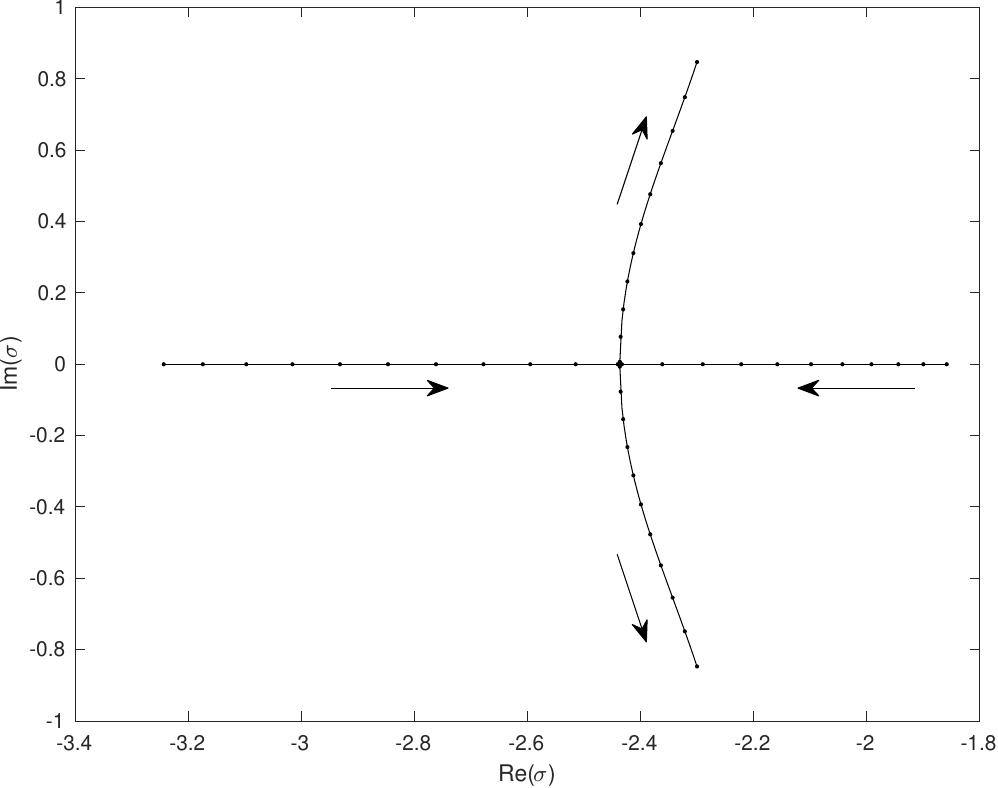}
    \caption{Numerically determined QNFs in a neighbourhood of the transition point at $h\approx 0.4645$, at which the two real QNFs with $l=0, \sigma(0) = -2, -3$ meet and branch into a conjugate pair of complex QNFs. Arrows indicate the direction of increasing $h$.\label{Plot3}}
\end{figure}

Figure \ref{Plot2} shows that pairs of QNFs do eventually meet and move off the real axis. In Figure \ref{Plot3} we show an example of one such interaction in the complex plane, which occurs when the quasinormal frequencies with $l=0, \sigma(0) = -2, -3$ coalesce and move into the complex plane at $h\approx 0.4645$. We note that (within the accuracy of the numerics) it appears that we cannot identify a smooth curve $\sigma(h)$ of QNFs passing through the point at which the QNFs meet (and hence cease to be simple). Whichever branch we pick, the curve will have to turn through an angle of $\pi/2$ as $h$ passes the critical value. The choices of $h$ to plot were determined by setting $h_i =0.4645 + \epsilon_i \abs{\epsilon_i}$, and taking $\epsilon_i$ to be spaced uniformly in $[-1,1]$. This figure was computed with a depth $k=3$ and $N=25$ gridpoints, see \S\ref{numerics}.

\begin{figure}
\begin{subfigure}{.5\textwidth}
  \centering
  \includegraphics[width=.9\linewidth]{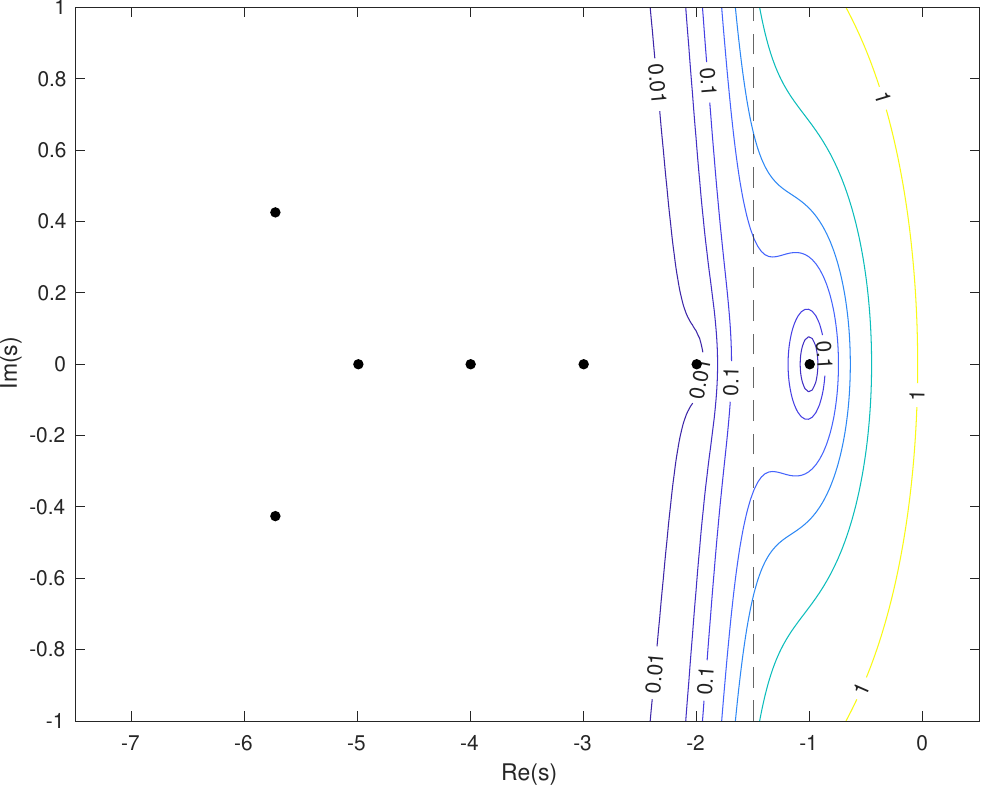}
  \caption{$k=1$}
  \label{Plot4:a}
\end{subfigure}%
\begin{subfigure}{.5\textwidth}
  \centering
  \includegraphics[width=.9\linewidth]{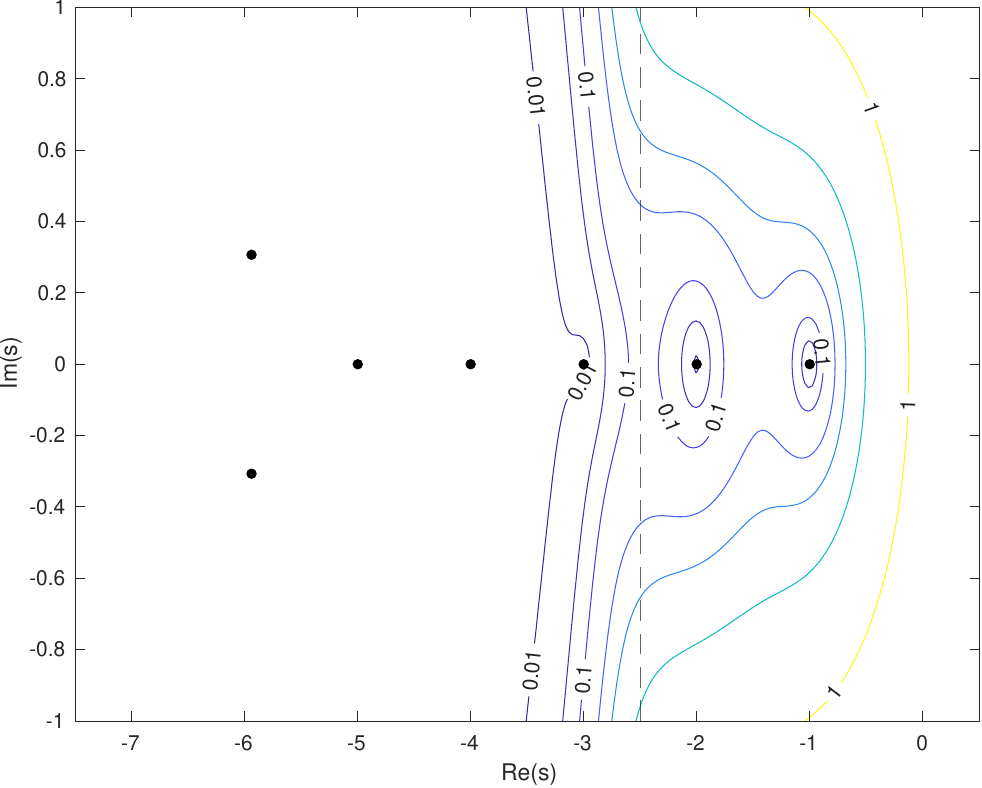}
  \caption{$k=2$}
  \label{Plot4:b}
\end{subfigure}%

\begin{subfigure}{.5\textwidth}
  \centering
  \includegraphics[width=.9\linewidth]{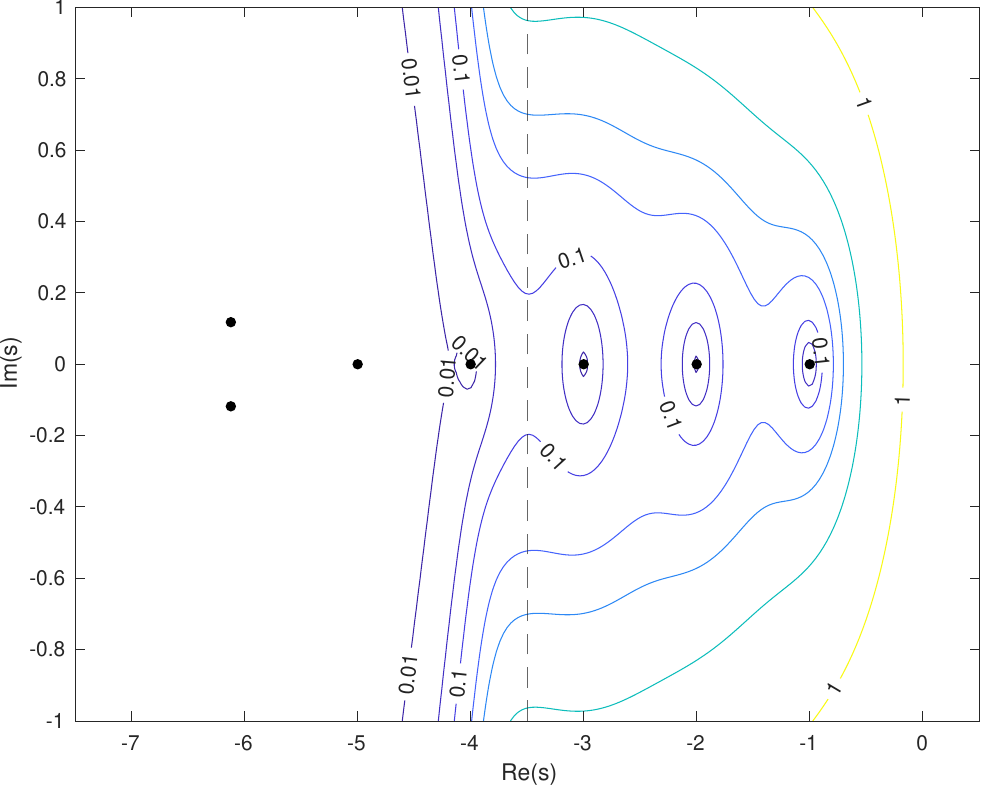}
  \caption{$k=3$}
  \label{Plot4:c}
\end{subfigure}%
\begin{subfigure}{.5\textwidth}
  \centering
  \includegraphics[width=.9\linewidth]{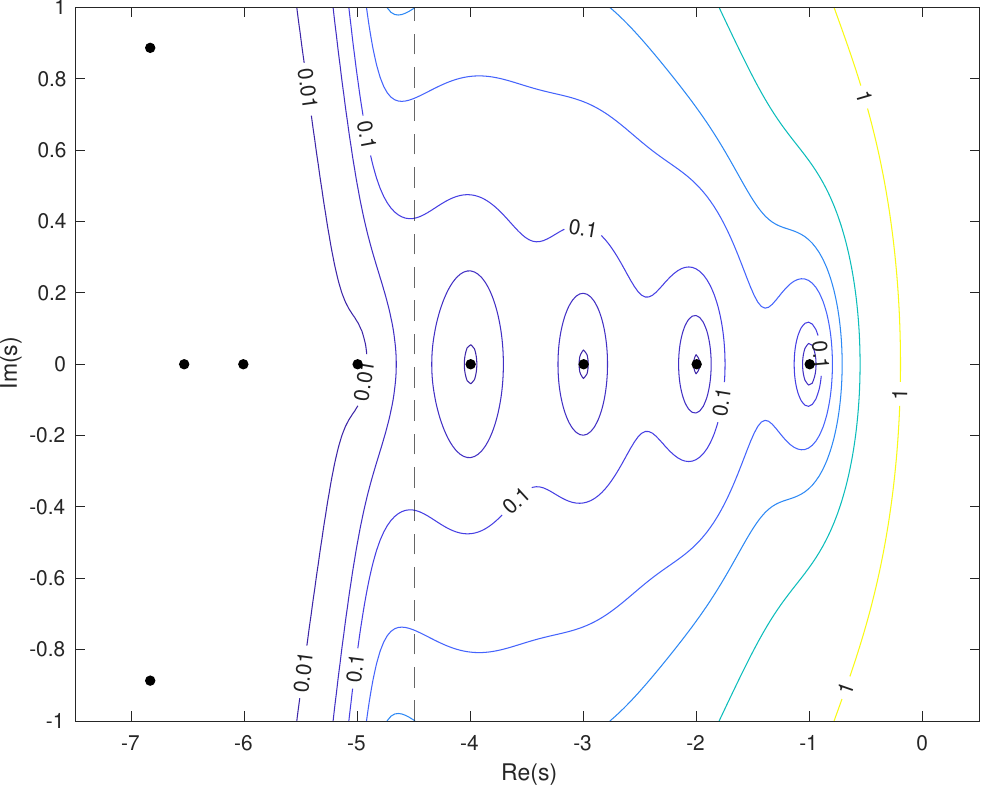}
  \caption{$k=4$}
  \label{Plot4:d}
\end{subfigure}%

\begin{subfigure}{.5\textwidth}
  \centering
  \includegraphics[width=.9\linewidth]{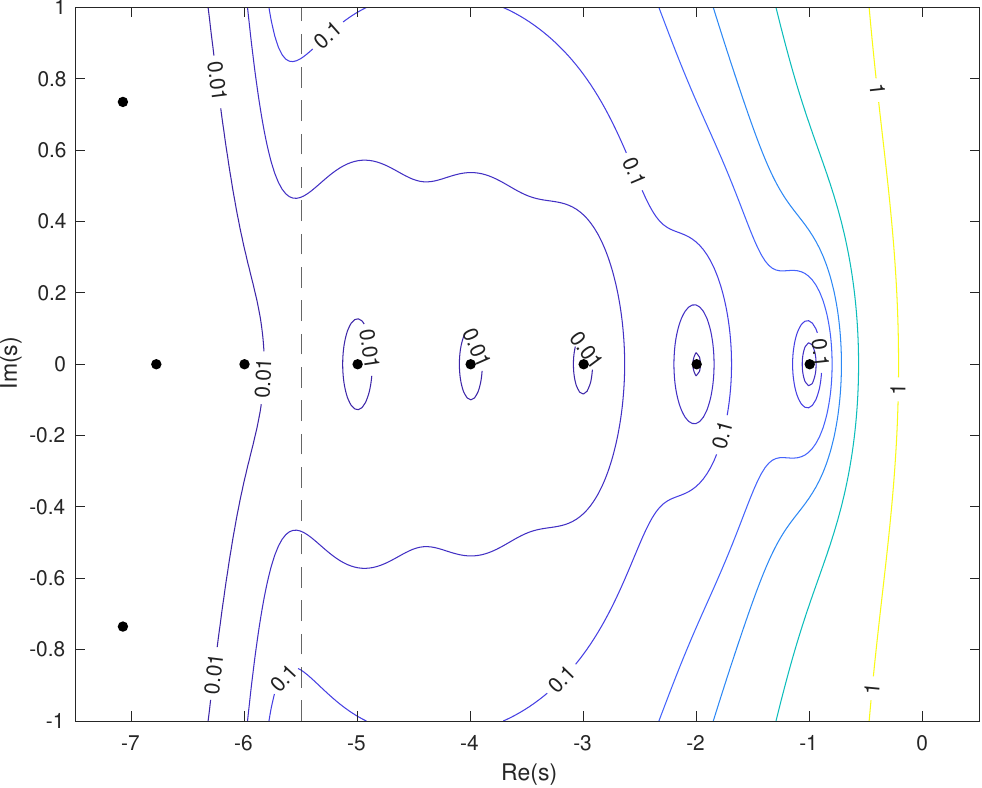}
  \caption{$k=5$}
  \label{Plot4:e}
\end{subfigure}%
\begin{subfigure}{.5\textwidth}
  \centering
  \includegraphics[width=.9\linewidth]{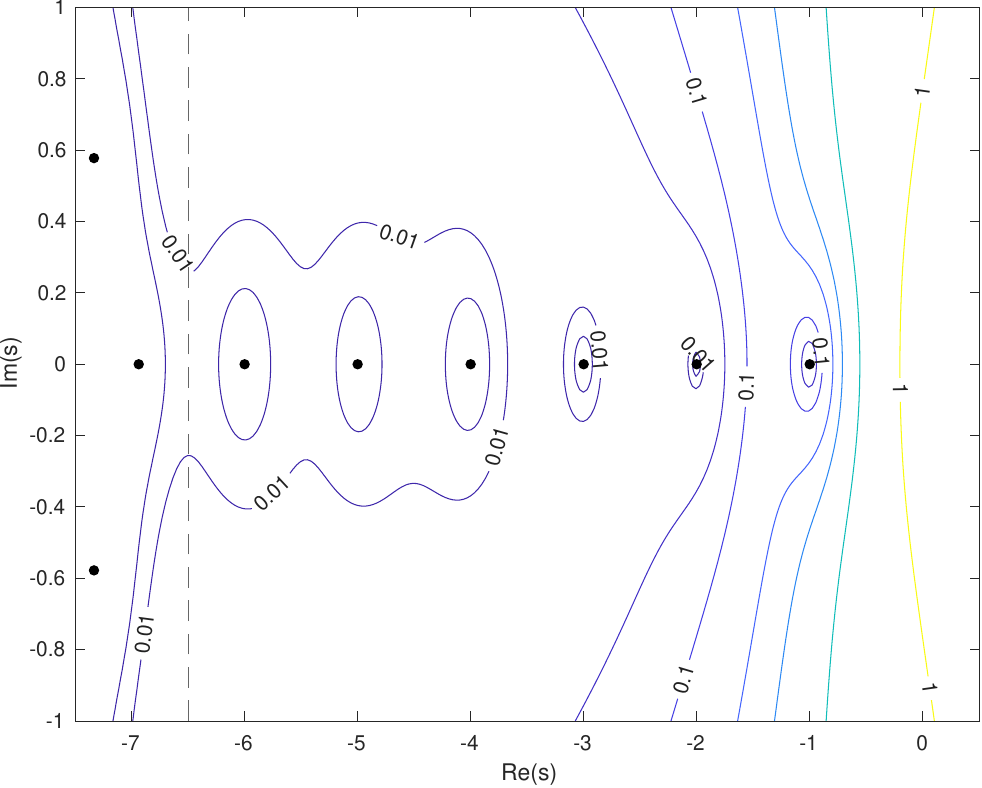}
  \caption{$k=6$}
  \label{Plot4:f}
\end{subfigure}
    \caption{Numerically computed contour lines $\norm{\hat{L}(s,0)^{-1}}{H^k \to H^k} = \epsilon^{-1}$ for $1\leqslant k \leqslant 6$. The black dashed line indicates the boundary of $U_k$. Black dots are the QNFs of $\hat{L}(s,0)^{-1}$ computed by the numerical algorithm with $N=35$.\label{Plot4}}
\end{figure}

\begin{figure}
\begin{subfigure}{.5\textwidth}
  \centering
  \includegraphics[width=.9\linewidth]{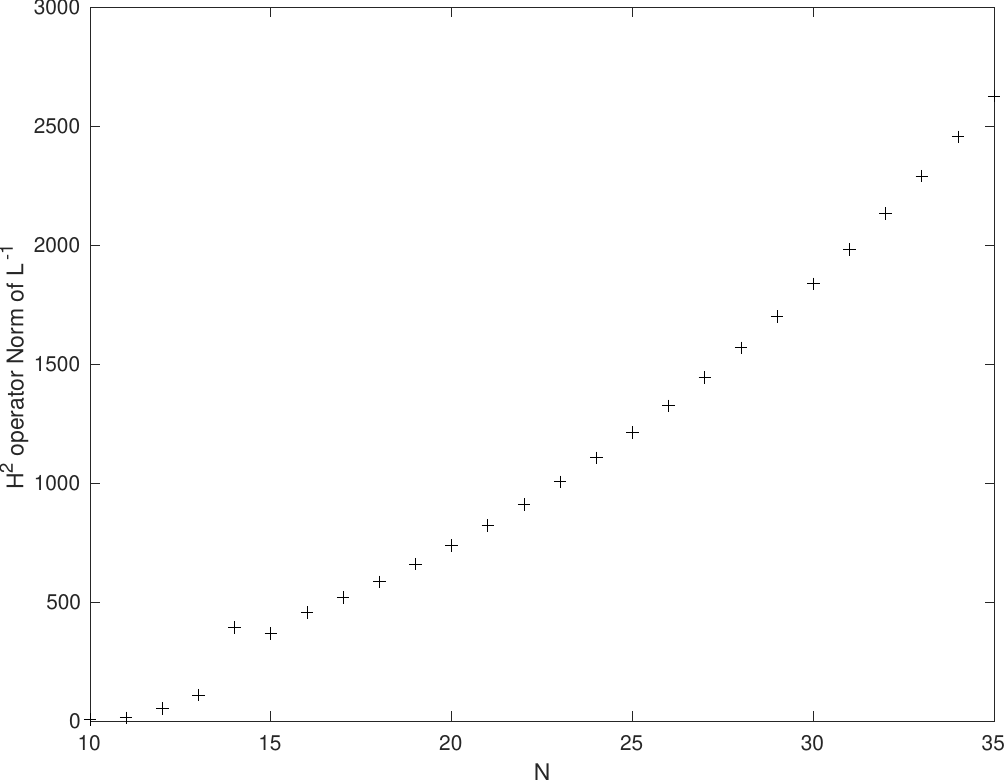}
  \caption{$k=2$}
  \label{Plot5:a}
\end{subfigure}%
\begin{subfigure}{.5\textwidth}
  \centering
  \includegraphics[width=.9\linewidth]{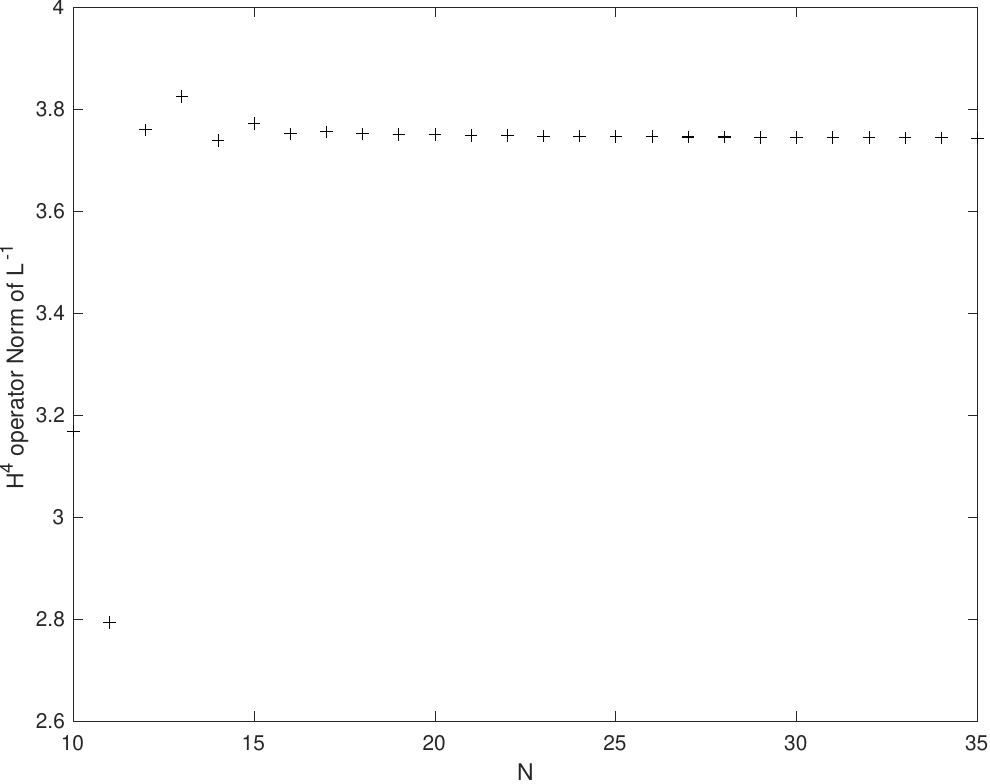}
  \caption{$k=4$}
  \label{Plot5:b}
\end{subfigure}
    \caption{Convergence of $\norm{\hat{L}^{-1}(s_0,0)}{H^k \to H^k}$ at $s_0 = -4+i$ for $k=2$ and $k=4$.\label{Plot5}}
\end{figure}

\subsection{The numerical scheme}\label{numerics}

The numerics in this section are performed using a null slicing, rather than the spacelike slicing introduced above, but the scheme can be readily adapted for a spacelike slicing. For convenience we will take $\kappa=1$ from now on. Setting
\[
\tau = t -  \log (1 +r)
\]
the metric takes the form
\begin{equation}
    g = -\left( 1- r^2 \right) d\tau^2 - 2  dr  d\tau + r^2 (d\theta^2 + \sin^2\theta d\phi^2).
\end{equation}
In order to find the quasinormal frequencies, we seek solutions to $\hat{L}(s, h)u = 0$ of the form
\[
u(r, \theta, \phi) = \frac{R(r)}{r} Y_{l, m}(\theta, \phi)
\]
If we define
\[
\mathcal{L}R:= \frac{d}{dr} \left( (1-r^2)\frac{dR}{dr}\right) - \frac{l(l+1)}{r^2} R + V_h R
\]
then 
\[
r\hat{L}(s, h)u = \mathcal{L}R - 2 s \frac{d R}{dr} 
\]
so that to find quasinormal frequencies, we are led to consider the solvability of
\begin{equation}\label{kEqn}
  \mathcal{L}R - 2 s \frac{d R}{dr} =f
\end{equation}
for given $f$, with $f(0) = R(0)=0$ and $R$ regular $r=1$. Rather than directly discretize \eqref{kEqn}, we first expand to a system of equations by differentiating the equation. We have the commutation relation
\[
\left[r \frac{d}{dr}, \mathcal{L} \right] = - 2 \mathcal{L} - 2 \left(r \frac{d}{dr}\right)^2- 2 r \frac{d}{dr}+ 2 V_h +  r \frac{dV_h }{dr}.
\]
Let $R^p = \left(r \frac{d}{dr} \right)^p R$,  $V_h^p = \left(r \frac{d}{dr} \right)^p V_h$ and $f^i = \left(r \frac{d}{dr} \right)^i f$. Then using the commutation relation we can show that \eqref{kEqn} implies
\begin{align*}
    \mathcal{L}R^p + \sum_{i=1}^{p+1} \alpha_i^p R^i + \sum_{i, j=0}^{p} \beta^{p}_{i,j} V_h^i R^j - s \sum_{i=0}^p \gamma^p_{i} \frac{d R^i}{dr} =\sum_{i=0}^p \mu^p_{i} f^i
\end{align*}
where $\alpha^p_i, \beta^p_{i,j}, \gamma^p_{i}$ are numerical (indeed integer) constants determined recursively by
\[
\alpha_i^{p+1} = \alpha_{i-1}^{p} + 2 \alpha_i^{p}\ \ (1\leqslant i \leqslant p), \qquad \alpha_{p+1}^{p+1} = \alpha_{p}^{p} + 2 \alpha_{p+1}^p-2, \qquad \alpha_{p+2}^{p+1} = \alpha_{p}^{p+1} -2
\]
with $\alpha_i^0 = 0$ for all $i$. Next we set $\beta^p_{i,j} = 0$ for all $i, j$ and recursively define
\[
\beta^{p+1}_{0,p} = \beta^{p}_{0,p-1} + 2 \beta^{p}_{0,p}+2, \qquad \beta^{p+1}_{1,p} = \beta^{p}_{0,p-1} + \beta^{p}_{1,p-1}+ 2 \beta^{p}_{1,p}+1.
\]
with
\[
\beta_{i,j}^{p+1} = \beta^p_{i-1, j}+ \beta^p_{i, j-1} + 2 \beta_{i, j} 
\]
otherwise. Finally, set $\gamma^0_0=1 = \mu^0_0$, $\gamma^0_i=0= \mu^0_i$ for all $i\neq 0$ and
\[
\gamma^{p+1}_i = \gamma^{p}_{i+1} +\gamma^{p}_i, \qquad \mu^{p+1}_i = \mu^p_{i-1} + 2 \mu^p_i.
\]

We can verify that $\alpha^p_{p+1} = -2 p$, which we expect as a consequence of the enhanced redshift effect (see \cite{Warnick2015OnHoles, Dafermos2013LecturesWaves}).

In order to construct our numerical scheme, we fix an integer $k\geqslant 0$, which we call the depth of the scheme. If \eqref{kEqn} holds, then the system of equations:
\begin{align}
    \left[\mathcal{L} -2 k r \frac{d}{dr} \right]R^p + 2 k R^{p+1}+  \sum_{i=1}^{p+1} \alpha_i^p R^i + \sum_{i, j=0}^{p} \beta^{p}_{i,j} V_h^i R^j - s \sum_{i=0}^p \gamma^p_{i} \frac{d R^i}{dr} &= \sum_{i=1}^p \mu^p_i f^i\ \ \ \ (0\leqslant p<k)\nonumber \\
\left[\mathcal{L} -2 k r \frac{d}{dr} \right]R^k +  \sum_{i=1}^{k} \alpha_i^k R^i + \sum_{i, j=0}^{k} \beta^{k}_{i,j} V_h^i R^j - s \sum_{i=0}^k \gamma^k_{i} \frac{d R^i}{dr}&= \sum_{i=1}^k \mu^k_i f^i \label{scheme}
\end{align}
will also hold. Here we have used the fact that $r \frac{d}{dr}R^p = R^{p+1}$ to arrange that we have the operator $\left[\mathcal{L} -2 k r \frac{d}{dr} \right]$ acting as the principle differential operator on all components. This is the approach taken to increase the working regularity in the analysis of \cite{Warnick2015OnHoles}. 

We now treat $R^0, \ldots, R^k$ as independent functions and we discretize on the interval $[0,1]$ using a pseudospectral method, following \cite{Trefethen2000SpectralMATLAB}. The constants $\alpha, \beta, \gamma$ are found recursively, and the derivatives $V_h^i$ are computed exactly using Matlab's \emph{Symbolic Math Toolbox} before discretisation. We note that $R^p(0) = 0$ which gives a Dirichlet boundary condition at one end of our interval, and we do not need a boundary condition at $r=1$ as the pseudospectral discretisation will impose smoothness there automatically.

After discretising on $N$ points, \eqref{scheme} becomes
\begin{equation}
    (A - sB)V = CF \label{genEV}
\end{equation}
for $(kN)\times (kN)$-matrices $A, B, C$ and column vectors $V, F$ which represent the discretisation of $(R^0, \ldots, R^k)$, $(f^0, \ldots, f^k)$ respectively. We work throughout at standard machine precision.

\subsubsection{The quasinormal spectrum}
If $\sigma$ is a quasinormal frequency, then we expect the generalised eigenvalue problem 
\[
(A - sB)V =0
\]
to have an eigenvalue near $\sigma$. Thus we can find the quasinormal frequencies by applying Matlab's generalised eigenvalue finder to \eqref{genEV}. However, by enlarging the original problem to a system we may have introduced spurious eigenvalues which correspond to vectors $V$ for which the condition 
\begin{equation}
    R^{p+1} = r \frac{dR^p}{dr}, \qquad 0\leqslant p <k \label{projcond}
\end{equation}
does not hold. To enforce this condition, we select only those eigenvalues of \eqref{genEV} for which (the discretised version of)
\[
\sum_{p=0}^{k-1} \norm{R^{p+1}- r \frac{dR^p}{dr}}{}^2 <e
\]
holds, where $e$ is a sufficiently small threshold parameter, which we take to be $10^{-1}$ for the computations in this paper. 

Plots \ref{Plot2:a}-\ref{Plot2:c} show the numerically computed quasinormal spectrum in the $l=0,1,2$ sector as $h$ varies, computed using $k=6, N=25$. Plot \ref{Plot2:d} shows the error in the scheme when computing the eigenvalue at $\sigma = -4$ for various values of $k$. We see (as has been observed in other situations \cite{Jaramillo2021PseudospectrumInstability}) that for a given value of $k$ the pseudospectral method in fact can accurately find quasinormal frequencies even outside the domain $U_k$ in which we expect the numerics to converge.

\subsubsection{The pseudospectra} In order to compute pseudospectra for different $k$, we need to numerically approximate $\norm{\hat{L}(s, 0)^{-1}}{H^k \to H^k}$. We can approximate this by computing
\[
\norm{\hat{L}(s, 0)^{-1}}{H^k \to H^k} \approx \norm{(A-sB)^{-1} C \Pi}{\ell^2 \to \ell^2}
\]
where $\Pi$ is the $L^2-$orthogonal projector onto the space of vectors $F$ of the form $(f^0, \ldots, f^k)$, where $f^i = \left(r \frac{d}{dr}\right)^i f$. This projection is necessary to account for the enlargement of our space by considering the system of higher derivatives. Since for such an $F$ we have\footnote{In fact this is the discretised version of a \emph{weighted} Sobolev norm, however since the weights only degenerate near $r=0$ this is adequate for our purposes.}
 $\norm{F}{\ell^2} \approx \norm{f}{H^k}$ we can approximately compute the $H^k$ operator norm of $\hat{L}(s, 0)^{-1}$ by the $\ell^2$ operator norm of the approximating matrix. 
 
 In Figure \ref{Plot4} we show the numerically computed pseudospectra for $k=1,\ldots, 6$. We see that in all cases the pseudospectrum is well behaved in the region $U_k$, but that the contours open out significantly once we leave this region. We expect that the fact that the contour curves can leave $U_k$ at all is a feature of the finite truncation. We observe the phenomenon noted above that the spectral method finds quasinormal frequencies accurately, even in the region of the plane that we expect significant numerical instability. For example in Figure \ref{Plot4:a} we see the first 5 frequencies accurately computed, even though only the first is actually in $U_1$.

In order to verify convergence of the numerical operator norm, in Figure \ref{Plot5} we show the approximated values of $\norm{\hat{L}(s, 0)^{-1}}{H^k \to H^k}$ at $s = -4+i$ for $k=2$ and $k=4$ as $N$ varies. As expected, in the $k=2$ case we see divergence, since for this $k$ our choice of $s$ does not belong to $U_k$. For the case $k=4$ we are in the region $U_k$, and we see good convergence. This can be compared to Figure 7 of \cite{Boyanov2024StructuralPseudospectrum}.

\section{Conclusion}

We have investigated the stability of the quasinormal spectrum of the conformal wave equation on the static patch of de Sitter. We find that the quasinormal frequencies are stable, provided the perturbing potential is small in a sufficiently high regularity norm. Conversely, one could instead interpret this as a spectral \emph{instability} for perturbing potentials which are not sufficiently regular at the cosmological horizon. We numerically verify our computations using a spectral method, and propose a definition for a family of pseudospectra that demonstrate good convergence properties and capture the (in)stability of the quasinormal frequencies.

\section*{Acknowledgements}
I am grateful to Jason Joykutty and Dejan Gajic for several valuable discussions on the topic of pseudospectral instability of black holes. I am especially grateful to Jos\'e Luis Jaramillo for discussions and very helpful comments on this manuscript. I am also grateful to the The Erwin Schr\"odinger International Institute for Mathematics and Physics in Vienna for hosting the programme ``Spectral Theory and Mathematical Relativity'' in Summer 2023, during which I started thinking about this problem.

\bibliographystyle{utphys} 
\bibliography{QNMreferences} 

\providecommand{\href}[2]{#2}\begingroup\raggedright\begin{thebibliography}{10}

\bibitem{Vasy2013MicrolocalDyatlov}
A.~Vasy, ``{Microlocal analysis of asymptotically hyperbolic and Kerr-de Sitter spaces (with an appendix by Semyon Dyatlov)},'' \href{http://dx.doi.org/10.1007/s00222-012-0446-8}{{\em Inventiones mathematicae} {\bfseries 194} no.~2, (11, 2013) 381--513}.

\bibitem{Warnick2015OnHoles}
C.~Warnick, ``{On Quasinormal Modes of Asymptotically Anti-de Sitter Black Holes},'' \href{http://dx.doi.org/10.1007/s00220-014-2171-1}{{\em Communications in Mathematical Physics} {\bfseries 333} no.~2, (2015) }.

\bibitem{Dyatlov2019MathematicalResonances}
S.~Dyatlov and M.~Zworski, ``{Mathematical Theory of Scattering Resonances},'' {\em Graduate Studies in Mathematics} (2019) . \url{https://api.semanticscholar.org/CorpusID:216573983}.

\bibitem{Gajic2021QuasinormalSpacetimes}
D.~Gajic and C.~Warnick, ``{Quasinormal Modes in Extremal Reissner–Nordstr{\"{o}}m Spacetimes},'' \href{http://dx.doi.org/10.1007/s00220-021-04137-4}{{\em Communications in Mathematical Physics} {\bfseries 385} no.~3, (2021) }.

\bibitem{Stucker2024QuasinormalHole}
T.~Stucker, ``{Quasinormal modes for the Kerr black hole},''.

\bibitem{Gajic2020AHoles}
D.~Gajic and C.~Warnick, ``{A model problem for quasinormal ringdown of asymptotically flat or extremal black holes},'' \href{http://dx.doi.org/10.1063/5.0024699}{{\em Journal of Mathematical Physics} {\bfseries 61} no.~10, (10, 2020) }.

\bibitem{Gajic2024QuasinormalSpacetimes}
D.~Gajic and C.~M. Warnick, ``{Quasinormal modes on Kerr spacetimes},''.

\bibitem{Aguirregabiria1996ScatteringApproach}
J.~Aguirregabiria and C.~Vishveshwara, ``{Scattering by black holes: a simulated potential approach},'' \href{http://dx.doi.org/10.1016/0375-9601(95)00937-X}{{\em Physics Letters A} {\bfseries 210} no.~4-5, (1, 1996) 251--254}.

\bibitem{Vishveshwara1996OnJourney}
C.~V. Vishveshwara, ``{On the black hole trail ...: A personal journey},'' in {\em 18th Conference of the Indian Association for General Relativity and Gravitation}, pp.~11--22.
\newblock 1996.

\bibitem{Nollert1996AboutHoles}
H.-P. Nollert, ``{About the significance of quasinormal modes of black holes},'' \href{http://dx.doi.org/10.1103/PhysRevD.53.4397}{{\em Phys. Rev. D} {\bfseries 53} (1996) 4397--4402}.

\bibitem{Nollert1999QuantifyingSystems}
H.-P. Nollert and R.~H. Price, ``{Quantifying excitations of quasinormal mode systems},'' \href{http://dx.doi.org/10.1063/1.532698}{{\em J. Math. Phys.} {\bfseries 40} (1999) 980--1010}.

\bibitem{Bizon2020AModes}
P.~Bizo{\'{n}}, T.~Chmaj, and P.~Mach, ``{A toy model of hyperboloidal approach to quasinormal modes},'' \href{http://dx.doi.org/10.5506/APhysPolB.51.1007}{{\em Acta Phys. Polon. B} {\bfseries 51} (2020) 1007}.

\bibitem{Ansorg2016SpectralSlices}
M.~Ansorg and R.~P. Macedo, ``{Spectral decomposition of black-hole perturbations on hyperboloidal slices},'' \href{http://dx.doi.org/10.1103/PhysRevD.93.124016}{{\em Physical Review D} {\bfseries 93} no.~12, (6, 2016) 124016}.

\bibitem{PanossoMacedo2018HyperboloidalCase}
R.~Panosso~Macedo, J.~L. Jaramillo, and M.~Ansorg, ``{Hyperboloidal slicing approach to quasi-normal mode expansions: the Reissner-Nordstr{\"{o}}m case},'' \href{http://dx.doi.org/10.1103/PhysRevD.98.124005}{{\em Phys. Rev. D} {\bfseries 98} no.~12, (2018) 124005}.

\bibitem{PanossoMacedo2020HyperboloidalSpacetime}
R.~Panosso~Macedo, ``{Hyperboloidal framework for the Kerr spacetime},'' \href{http://dx.doi.org/10.1088/1361-6382/ab6e3e}{{\em Class. Quant. Grav.} {\bfseries 37} no.~6, (2020) 65019}.

\bibitem{Jaramillo2021PseudospectrumInstability}
J.~L. Jaramillo, R.~Panosso~Macedo, and L.~Al~Sheikh, ``{Pseudospectrum and Black Hole Quasinormal Mode Instability},'' \href{http://dx.doi.org/10.1103/PhysRevX.11.031003}{{\em Phys. Rev. X} {\bfseries 11} no.~3, (2021) 31003}.

\bibitem{AlSheikh2022ScatteringSystems}
L.~Al~Sheikh, {\em {Scattering resonances and Pseudospectrum: stability and completeness aspects in optical and gravitational systems}}.
\newblock PhD thesis, Universit{\'{e}} Bourgogne Franche-Comt{\'{e}}, 2022.

\bibitem{Cheung2022DestabilizingFlea}
M.~H.-Y. Cheung, K.~Destounis, R.~P. Macedo, E.~Berti, and V.~Cardoso, ``{Destabilizing the Fundamental Mode of Black Holes: The Elephant and the Flea},'' \href{http://dx.doi.org/10.1103/PhysRevLett.128.111103}{{\em Physical Review Letters} {\bfseries 128} no.~11, (3, 2022) 111103}.

\bibitem{Sarkar2023PerturbingConstant}
S.~Sarkar, M.~Rahman, and S.~Chakraborty, ``{Perturbing the perturbed: Stability of quasinormal modes in presence of a positive cosmological constant},'' \href{http://dx.doi.org/10.1103/PhysRevD.108.104002}{{\em Physical Review D} {\bfseries 108} no.~10, (11, 2023) 104002}.

\bibitem{Arean2023PseudospectraModes}
D.~Are{\'{a}}n, D.~G. Fari{\~{n}}a, and K.~Landsteiner, ``{Pseudospectra of holographic quasinormal modes},'' \href{http://dx.doi.org/10.1007/JHEP12(2023)187}{{\em Journal of High Energy Physics} {\bfseries 2023} no.~12, (12, 2023) 187}.

\bibitem{Destounis2023Black-holePseudospectrum}
K.~Destounis and F.~Duque, ``{Black-hole spectroscopy: quasinormal modes, ringdown stability and the pseudospectrum},''.

\bibitem{Cownden2024TheAdS}
B.~Cownden, C.~Pantelidou, and M.~Zilh{\~{a}}o, ``{The pseudospectra of black holes in AdS},'' \href{http://dx.doi.org/10.1007/JHEP05(2024)202}{{\em Journal of High Energy Physics} {\bfseries 2024} no.~5, (5, 2024) 202}.

\bibitem{Boyanov2024StructuralPseudospectrum}
V.~Boyanov, V.~Cardoso, K.~Destounis, J.~L. Jaramillo, and R.~Panosso~Macedo, ``{Structural aspects of the anti-de Sitter black hole pseudospectrum},'' \href{http://dx.doi.org/10.1103/PhysRevD.109.064068}{{\em Phys. Rev. D} {\bfseries 109} no.~6, (2024) 64068}.

\bibitem{Jaramillo2022PseudospectrumTransientsb}
J.~L. Jaramillo, ``{Pseudospectrum and binary black hole merger transients},'' \href{http://dx.doi.org/10.1088/1361-6382/ac8ddc}{{\em Classical and Quantum Gravity} {\bfseries 39} no.~21, (11, 2022) 217002}.

\bibitem{Gasperin2022EnergyProduct}
E.~Gasper{\'{i}}n and J.~L. Jaramillo, ``{Energy scales and black hole pseudospectra: the structural role of the scalar product},'' \href{http://dx.doi.org/10.1088/1361-6382/ac5054}{{\em Classical and Quantum Gravity} {\bfseries 39} no.~11, (6, 2022) 115010}.

\bibitem{BessonBlackPreparation}
J.~Besson, V.~Boyanov, and J.~L. Jaramillo, ``{Black hole quasi-normal modes as eigenvalues: definition and stability problem, in preparation},''.

\bibitem{Hintz2021QuasinormalSpace}
P.~Hintz and Y.~Xie, ``{Quasinormal modes and dual resonant states on de Sitter space},'' \href{http://dx.doi.org/10.1103/PhysRevD.104.064037}{{\em Phys. Rev. D} {\bfseries 104} no.~6, (2021) 64037}.

\bibitem{Hintz2022QuasinormalHoles}
P.~Hintz and Y.~Xie, ``{Quasinormal modes of small Schwarzschild–de Sitter black holes},'' \href{http://dx.doi.org/10.1063/5.0062985}{{\em Journal of Mathematical Physics} {\bfseries 63} no.~1, (1, 2022) }.

\bibitem{Joykutty2022ExistenceHoles}
J.~Joykutty, ``{Existence of Zero-Damped Quasinormal Frequencies for Nearly Extremal Black Holes},'' \href{http://dx.doi.org/10.1007/s00023-022-01202-z}{{\em Annales Henri Poincar{\'{e}}} {\bfseries 23} no.~12, (12, 2022) 4343--4390}.

\bibitem{Trefethen1999SpectraPseudospectra}
L.~N. Trefethen, \href{http://dx.doi.org/10.1007/978-3-662-03972-4{\_}6}{``{Spectra and Pseudospectra},''} in {\em The Graduate Student's Guide to Numerical Analysis '98: Lecture Notes from the VIII EPSRC Summer School in Numerical Analysis}, M.~Ainsworth, J.~Levesley, and M.~Marletta, eds., pp.~217--250.
\newblock Springer Berlin Heidelberg, Berlin, Heidelberg, 1999.
\newblock \url{https://doi.org/10.1007/978-3-662-03972-4_6}.

\bibitem{EmbreePseudospectraGateway}
M.~Embree and L.~N. Trefethen, {\em {Pseudospectra Gateway}}.
\newblock http://www.comlab.ox.ac.uk/pseudospectra.

\bibitem{vanDorsselaer1993LinearProblems}
J.~van Dorsselaer, J.~Kraaijevanger, and M.~Spijker, ``{Linear stability analysis in the numerical solution of initial value problems},'' \href{http://dx.doi.org/10.1017/S0962492900002361}{{\em Acta Numerica} {\bfseries 2} (1, 1993) 199--237}.

\bibitem{Joykutty2024QuasinormalHoles}
J.~Joykutty, \href{http://dx.doi.org/10.17863/CAM.106739}{{\em {Quasinormal Modes of Nearly Extremal Black Holes}}}.
\newblock PhD thesis, Department of Applied Mathematics And Theoretical Physics, Cambridge U., 2024.

\bibitem{Dafermos2013LecturesWaves}
M.~Dafermos and I.~Rodnianski, ``{Lectures on black holes and linear waves},'' {\em Clay Math. Proc.} {\bfseries 17} (2013) 97--205.

\bibitem{Trefethen2000SpectralMATLAB}
L.~N. Trefethen, \href{http://dx.doi.org/10.1137/1.9780898719598}{{\em {Spectral Methods in MATLAB}}}.
\newblock Society for Industrial and Applied Mathematics, 1, 2000.

\end{thebibliography}\endgroup

\end{document}